\documentclass[aps, pra,preprint,superscriptaddress,amsmath,showpacs]{revtex4-1}

\usepackage[T1]{fontenc}
\usepackage[utf8]{inputenc}
\usepackage{rotating}
\usepackage{bm}        
\usepackage{amssymb}   
\usepackage{amsmath}
\usepackage{bm}
\usepackage[capitalise]{cleveref}
\usepackage{amsmath,amssymb,amsfonts, bm,amsthm} 
\usepackage{graphicx} 
\usepackage{color}
\newcommand{\beq}{\begin{equation}}
\newcommand{\eeq}{\end{equation}}
\newcommand{\bra}[1]{\langle #1 |}
\newcommand{\ket}[1]{| #1 \rangle}

\newcommand{\beqs}{\begin{subequations}}
\newcommand{\eeqs}{\end{subequations}}
\newcommand{\nbar}{\bar{n}}

\newtheorem{mythm}{Theorem}

\newcommand{\gbar}{\bar\gamma}

\newcommand{\Real}{\operatorname{Re}}
\newcommand{\Imag}{\operatorname{Im}}
\newcommand{\akl}{\hat{a}_{\bm{k}\lambda}}
\newcommand{\akld}{\akl^\dagger}
\newcommand{\ketbra}[2]{\ket{#1}\bra{#2}}
\newcommand{\kl}{\bm{k}\lambda}
\newcommand{\klp}{\bm{k'}\lambda'}
\newcommand{\half}{\frac{1}{2}}
\newcommand{\Vsb}{\hat{V}_{SR}^I}
\DeclareMathOperator{\Tr}{Tr_R}
\DeclareMathOperator{\Tre}{Tr_E}

\begin{document}
\date{\today}
\flushbottom
\title{Coherent dynamics of V-type systems driven by time-dependent incoherent radiation} 
\author{Amro Dodin}
\affiliation{Chemical Physics Theory Group, Department of Chemistry, and Center for Quantum Information and Quantum Control, University of Toronto, Toronto, Ontario, M5S 3H6, Canada}
\author{Timur V. Tscherbul}
\affiliation{Department of Physics, University of Nevada, Reno, NV, 89557, USA}
\author{Paul Brumer}
\affiliation{Chemical Physics Theory Group, Department of Chemistry, and Center for Quantum Information and Quantum Control, University of Toronto, Toronto, Ontario, M5S 3H6, Canada}

\begin{abstract}
  Light induced processes in nature occur by irradiation with slowly turned-on incoherent light. The general case of time-dependent incoherent excitation is solved here analytically for V-type systems using a newly developed master equation method. Clear evidence emerges for the disappearance of radiatively induced coherence as turn-on times of the radiation exceed characteristic system times. The latter is the case, in nature, for all relevant dynamical time scales for other than nearly degenerate energy levels. We estimate that, in the absence of non-radiative relaxation and decoherence, turn-on times slower than 1 ms (still short by natural standards) induce Fano coherences between energy eigenstates that are separated by less than 0.9 cm$^{-1}$.
\end{abstract}

\maketitle
\clearpage
\newpage

\section{Introduction}
\label{sec:Intro}
A number of femtosecond laser spectroscopy studies \cite{tiwari_electronic_2013,collini_coherently_2010,engel_evidence_2007,hau_light_1999} on components of light harvesting systems, such as the Fenna-Matthews-Olson (FMO) or PC645 photosynthetic complexes, show that irradiation with fs pulses results in coherent molecular energy transfer dynamics. 
These observations have been interpreted as demonstrating a role for quantum coherent dynamics in biological systems. 
However, as has been repeatedly argued \cite{grinev_realistic_2015,tscherbul_partial_2015,chenu_thermal_2015,tscherbul_long-lived_2014,sadeq_transient_2014,olsina_can_2014,tscherbul_excitation_2014,brumer_molecular_2012,pachon_computational_2012,mancal_exciton_2010,ou_coherence_2008,agarwal_quantum_2001,jiang_creation_1991} both formally and computationally, the response of a molecule to coherent laser light is dramatically different from that to natural incoherent radiation, such as sunlight. 
For example, in the absence of a decohering environment the pulsed laser case shows persistent molecular coherences, whereas the incoherent case yields a complex mixture of molecular energy eigenstates \cite{tscherbul_long-lived_2014,dodin_quantum_2016}.
These results cast doubt on the relevance of the experimentally observed molecular coherences to natural light induced and light harvesting processes.

Such studies, barring one \cite{grinev_realistic_2015} have all relied upon the sudden turn-on of the radiation, which is both unnatural and which generates initial coherence due to the abrupt turn-on of the light.
That is, they focused on the fate of coherences after they were generated by sudden turn-on of the radiation.
However, natural turn-on of light (e.g. sunrise for photosynthesis,
 or the blinking of an eye for vision) is very slow by comparison with molecular time scales, motivating further studies of the time evolution of systems subject to time dependent incoherent excitation, and of the associated Fano coherences discussed below.
This study is carried out here on the generic V-system, analytically exposing the dependence of the system evolution and the associated coherences on the turn-on time.
The results clearly show that the slower the turn-on time, the less the generated molecular coherences.
In particular, with natural turn-on times, no molecular coherences will appear between other than near-degenerate levels.
For example, with turn-on times on the order of 1 ms, which is still very short compared to natural turn-on times, coherences will be established only between levels spaced by 0.9 cm$^{-1}$.
A one second turn-on will only induce coherences in levels spaced by $9\times 10^{-4}$ cm$^{-1}$.

To consider such coherent effects rigorously, we examine the most general picture of weak-field incoherent light-matter interactions.
This is given by the Bloch-Redfield (BR) master equations, in which the populations and coherences of a reduced density matrix are treated on an equal footing \cite{cohen-tannoudji_atom-photon_1992}.
The Pauli rate law equations underlying, for example, the Einstein theory for excitation by incoherent light \cite{loudon_quantum_2000}, can be obtained from the BR theory by neglecting the non-secular terms that couple populations and coherences.
However, these non-secular terms are responsible for Fano interference between different incoherent excitation pathways \cite{fleischhauer_lasing_1992,kozlov_inducing_2006,agarwal_quantum_1974}
The existence of Fano coherences in incoherently driven systems has sparked considerable interest in the context of naturally occurring LHC's and artificial photovoltaics \cite{fleischhauer_lasing_1992,kozlov_inducing_2006,kiffner_progress_2010} where it has been proposed as a mechanism for enhancing the efficiency of quantum heat engines \cite{dorfman_photosynthetic_2013,scully_quantum_2011}.

The Fano coherences differ significantly in origin from the coherences induced by coherent light. 
As shown below, they can be understood most easily in terms of a number state picture where absorption of light with frequency $\omega=\omega_i$ leads a system in the ground state $\ket{g}$ to make a transition to the excited statesx $\ket{e_i}$ and gain phase according to the complex phase of the corresponding transition dipole moment \cite{dodin_quantum_2016}. 
For simplicity assume that the transition dipole moments are real and positive.
The Fano coherences arise due to simultaneous excitation from the ground state to both excited states, producing a coherent in-phase superposition of the excited states.
In contrast, the coherences arising from excitation of the system with coherent sources are a consequence of the phase relations between the transition frequencies. 
In the number state (photon) picture, coherent light is given by a coherent superposition of number states which contains the phase information as coherences between the number states at the corresponding frequencies.
The radiation field coherences can then be ``transferred'' to the system through the dipole interaction.

Our previous work has explored the role of the Fano coherences in the dynamical evolution of the V-system in the weak pumping limit \cite{tscherbul_long-lived_2014,dodin_quantum_2016}.
We have derived the Bloch-Redfield equations for a general class of multilevel systems \cite{tscherbul_partial_2015} and identified the parameter dependence of the dynamical evolution of the V-system \cite{dodin_quantum_2016}.
Here, these studies are significantly extended by considering the regime of non-stationary time-dependent incoherent radiation.
Section II develops a model for a time-dependent field which is then used to generalize the Bloch-Redfield equations to the case of time-dependent fields in \Cref{sec:QME}.
\Cref{sec:Vsys} uses the BR equations to consider a weakly pumped V-system and presents the general analytical solution, which is examined more closely for the limiting cases of a closely spaced $\Delta\ll\gamma$ system in \Cref{sec:OD} and a system with wide level spacing $\Delta\gg\gamma$ in \Cref{sec:UD}, where $\Delta$ is the excited state splitting and $\gamma$ is the rate of spontaneous emission.
Finally, \Cref{sec:Conc} summarizes our results.

Note that this paper deals with the molecular coherences generated by the incident light.
Those associated with, for example, donor excitation in a donor-acceptor system, briefly discussed in \cite{grinev_realistic_2015}, will be discussed in detail elsewhere \cite{tscherbul_donor-acceptor_????,dodin_slow_????}. Further, these studies do not include, but do motivate including, a second bath that would 
model, e.g., a bosonic environment. In that case, where the system
is coupled to two baths, the long time steady state is a 
"transport problem", with flow of energy from the radiation field
to the second bath. Studies of this kind, which would extend 
work such as that in Ref. \cite{pachonpre}, are in progress.

\section{Modelling the Time-Dependent Field}
\label{sec:Field}

\subsection{Properties of Field Dynamics}
\label{sec:DynProp}
In this section, we first consider   a model for a time varying radiation field corresponding to the slow turn on of incoherent light (e.g., a thermal field attenuated by a variable filter),
 that is an isotropic  unpolarized radiation field with  constant frequency distribution but time varying intensity. Furthermore, let the radiation field be diagonal in the number state representation at all times.

The isotropy and unpolarized property is straightforward to implement through the following relationship, enforced at all times $t$ and for all non-negative integers $m$:
\beq
{\langle \hat{n}_{\bm{j}\lambda}^m\rangle (t)}={\langle \hat{n}_{\bm{k}\mu}^m\rangle(t)} \mbox{; if }|\bm{j}|=|\bm{k}|
\label{eq:Isotropy}
\eeq
where $\hat{n}_{\bm{j}\lambda}$ is the number operator for the field mode with wave vector $\bm{j}$ and linear polarization $\lambda=1,2$ and $\langle \hat{A} \rangle (t)$ is the expectation value of the field operator $\hat{A}$.
That is, \cref{eq:Isotropy} states that all statistical moments, $m$, of the field mode depend only on the magnitude of the wave vector and not on its direction or polarization.
Therefore, the statistical distribution of all field modes with the same wave vector magnitude is identical, realizing an isotropic and unpolarized field.

The requirement for a time-independent frequency distribution is equivalent to assuming that the temperature of a blackbody source does not change. That is,
\beq
\frac{\langle \hat{n}_{\bm{j}\lambda}\rangle (t)}{\langle \hat{n}_{\bm{k}\mu}\rangle(t)}=Constant \mbox{ for all } t>0
\label{eq:Freq}
\eeq
This leads to a source with time-independent bandwidth, $\Delta \omega$, since the bandwidth is a property of the frequency distribution.
Consequently, this leads to a time-independent coherence time $\tau_c$ for all $t>0$ since $\tau_c \Delta\omega \sim 1$ \cite{scully_quantum_1997,mandel_optical_1995}.

The time-dependence of the field is characterized by the varying intensity of the incident light.
Consider the intensity operator of a multimode field:
\beq
\hat{I}= \bm{\hat{E}}^{(-)} \cdot \bm{\hat{E}}^{(+)} = \left(\sum_{\bm{k} \lambda} \bm{\epsilon}_{\bm{k}\lambda} \xi_k \hat{a}_{\bm{k}\lambda} e^{i\nu_k t}\right) \cdot \left(\sum_{\bm{j}\mu} \bm{\epsilon}_{\bm{j}\mu} \xi_j \hat{a}^\dagger_{\bm{j}\mu} e^{-i\nu_j t}\right)
\label{eq:Int}
\eeq
For a given field mode with wavevector $\bm{k}$ and linear polarization $\lambda=1,2$, $\hat{a}_{\bm{k}\lambda}$ and $\hat{a}_{\bm{k}\lambda}^\dagger$ are the annihilation and creation operators, $\bm{\epsilon}_{\bm{k},\lambda}$ is the corresponding polarization vector, $\nu_k$ is the frequency of the field mode and $\xi_k=(\hbar \nu_k/2\epsilon_0V_{ph})^{1/2}$ is the electric field per photon, where $V_{ph}$ is the photon volume.  Here,
$\bm{\hat{E}}^{(\pm)}$ are the positive and negative frequency components of the electric field.
For an unpolarized isotropic field, \cref{eq:Int} reduces to
\beq
\hat{I}=\sum_k2|\xi_k|^2\left(\hat{n}_{k}+\frac{1}{2}\right)
\label{eq:IsoInt}
\eeq
where $\hat{n}_{k}=\sum_\lambda\sum_{\bm{j}:|j|=|k|}\hat{n}_{\bm{j}\lambda}$ is the total occupation number of field modes with wave vector magnitude $k$.
The only part of \cref{eq:IsoInt} that depends on the field properties is the total occupation number operator.
Therefore, to obtain a time-dependent intensity expectation value, the radiation field density matrix must change over time such that
\beq
n_{\bm{k}\lambda}(t)=\langle \hat{n}_{\bm{k}\lambda}\rangle(t)=\Tr\left\{\hat{n}_{\bm{k}\lambda}\hat{\rho}_R(t)\right\}=n_{\bm{k}\lambda}f(t)
\label{eq:nt}
\eeq
where $\Tr$ is the trace over the radiation field.
The turn on function must be identical for all modes due to the restrictions from \cref{eq:Isotropy,eq:Freq}.
For computational simplicity we consider a slow turn on envelope of the form
\beq
f(t)=1-e^{-\alpha t}
\label{eq:ft}
\eeq
where $\alpha$ is a constant characterizing the turn on rate, with a corresponding turn on time scale $\tau_r=1/\alpha$.
Appendix \ref{sec:Generality} outlines the generalization of the results  for arbitrary turn on functions through their expansion in a Laplace-like basis.

Finally, Eq. (\ref{eq:nt}) gives a source that is quasi-canonical in that no coherences exist between number states of the bath.
Intuitively, this corresponds to a bath that is similar to the incoherent fields studied previously \cite{dodin_quantum_2016} (e.g. Blackbody radiation field) at each instant in time.

\subsection{Realizing the Time-Dependent Field}
Since our interest is in the system that is irradiated by the incoherent light, the latter acts as an time-dependent external bath that is coupled to the system.
Treating a time-dependent bath within the standard framework of open quantum systems poses  challenges that are not present in the stationary field case.
That is, in standard density matrix theory the System-Bath composite is assumed to be closed, allowing for the development of the density matrix theory in terms of the Unitary Hamiltonian evolution of the total state vector in the joint Hilbert Space of the system and bath \cite{breuer_theory_2007}.
However, the time evolution of the radiation field described in Sec. IIA does not arise through the typical Hamiltonian evolution of a system  $+$ bath.

\begin{figure}[ht]
	\centering
	\includegraphics[width=\textwidth, trim = 0 0 0 0]{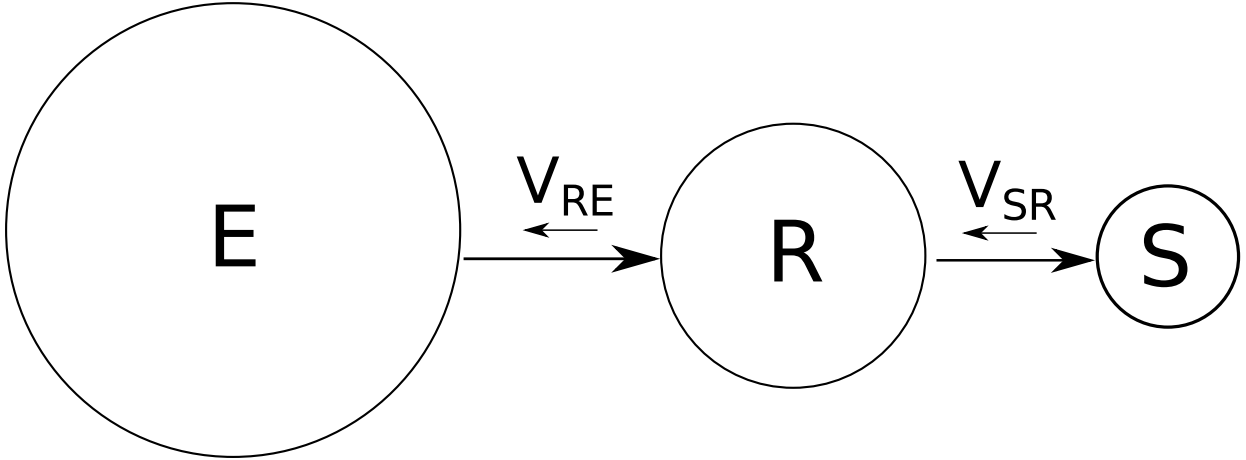}
	\renewcommand{\figurename}{Fig.}
	\caption{A schematic representation of the coupling between the molecular system, S, radiation bath, R, and environment, E. The system and environment are not directly coupled. The system-bath interaction, $V_{SR}$, and bath-environment interaction, $V_{RE}$ satisfy the Born-Markov approximation.}
\label{fig:EnvSchem}
\end{figure}

Instead,  we need to consider an additional environment $E$ coupled to the radiation bath, but not to the system, in the Born-Markov approximation, and let the radiative bath be coupled to the system in the Born-Markov regime.
This set up is sketched in \cref{fig:EnvSchem}.
Intuitively, the environment corresponds to the physical system that produces the dynamics of the field on the system.
This hierarchical approach has the benefit of allowing  the use of the standard approach to open quantum systems \cite{breuer_theory_2007,blum_density_2012}, since the system-bath-environment composite is closed and hence evolves according to unitary Hamiltonian dynamics.
To obtain well posed problems for the dynamics of the bath and the system,  the system-bath-environment is assumed to be  initially in a separable state $\hat{\rho}_0=\hat\rho_S\otimes\hat\rho_R\otimes\hat\rho_E$.

The Born approximation to the system-bath coupling implies that the influence of the system on the bath dynamics  is negligible. 
Therefore, we can treat the environment-bath composite using the standard approach to computing the reduced dynamics of the bath.
We can select the environment and interaction potential such that the dynamics of the bath correspond to that discussed in Sec. IIA.
In the resultant picture, the properties of the bath operators can be treated by exact analogy to the system operators in a typical system-bath case. For example, the two time correlation function for two Schr\"odinger Operators, $\hat{A}$ and $\hat{B}$, on the bath Hilbert Space, $\mathcal{H}_R$ is given by:
\beq
\langle \hat{A}(s)\hat{B}(t) \rangle = \Tr\left\{\hat{A} \hat{U}(s,t) \hat{B}\hat{U}(t,0) \hat\rho_R(0)\right\}
\label{eq:BathCorr}
\eeq
where $\hat\rho_R(0)$ is the initial state of the bath, assumed to be the vacuum state in our case, and $\hat{U}(s,t)$ is the bath propagation operator from time $s$ to time $t$.
After obtaining the bath dynamics, the resulting time-varying density matrix can be used to determine the dynamics of the system while neglecting any further role of the environment, E.

\section{Derivation of the Bloch-Redfield Equations for Time-Dependent Baths}
\label{sec:QME}
Traditional master equations assume a time independent bath.  
Therefore, to treat turn-on effects, we derive a generalization to the time dependent bath.
Consider a multilevel system interacting with a quantized incoherent radiation field, as described in Sec. II, under the dipole and rotating wave approximations.
Such a system is characterized by the total Hamiltonian
\beq
\hat{H}_T=\hat{H}_S + \hat{H}_R + \hat{H}_E + \hat{V}_{SR} +\hat{V}_{RE}
\label{eq:Htot}
\eeq
where $\hat{H}_S=\sum_i E_i\ket{i}\bra{i}$ is the system Hamiltonian, $\hat{H}_R=\sum_{\bm{k}\lambda}\hbar\nu_k\akl\akld$ is the radiation bath Hamiltonian and $\hat{H}_E$ is the environment Hamiltonian. The operator $\hat{V}_{RE}$ is the bath-environment interaction potential. The bosonic creation and annihilation operators of the field mode with wavevector $\bm{k}$, frequency $\nu_k$ and polarization $\lambda =1,2$ are given by $\akld$ and $\akl$ respectively. $\hat{V}_{SR}$ is the system-bath interaction potential given by \cite{scully_quantum_1997}
\beq
\hat{V}_{SR}=-\bm{\hat{\mu}} \cdot\sum_{\bm{k},\lambda}\left(\frac{\hbar \nu_k}{2\epsilon_0 V}\right)^{1/2}\bm{\epsilon}_{\bm{k}\lambda}(\hat{a}_{\bm{k}\lambda}-\hat{a}^\dagger_{\bm{k}\lambda})
\label{eq:Vint}
\eeq
where $V$ is the quantization volume, $\bm{\hat{\mu}}$ is the dipole moment operator of the system and $\bm{\epsilon}_{\kl}$ is the polarization vector of the field mode with wavevector $\bm{k}$ and polarization $\lambda$.

The bath-environment interaction potential and environment Hamiltonian are not specified, but are chosen to produce the bath dynamics discussed in Sec. II.
Under the Born-Markov approximation, the system does not contribute to the bath evolution, induced by its interaction with the environment.
In other words, the effects of $\Vsb$ and $\hat{H}_S$ on the dynamics of the radiation field + environment can be neglected.
This produces a density operator, $\hat{\rho}_{RE}$, on the radiation field + environment Hilbert space such that the radiation field density matrix $\hat{\rho}_R (t)=\Tre\hat{\rho}_{RE}$ follows the conditions described in Sec. II, where $\Tre$ is the trace over the environment $E$.

Transforming \cref{eq:Vint} into the interaction picture gives
\beq
\hat{V}_{SR}^I(t)= \hbar\sum_{i\leq j}\sum_{\kl} g_{\kl}^{(i,j)}(\akl \ketbra{j}{i}e^{i(\omega_{ij}-\nu_k)t}+\mbox{H.c.})
\label{eq:VSB}
\eeq
where $g_{\kl}^{(i,j)}=\left(\frac{\hbar\nu_k}{2 \epsilon_0 V}\right)^{\frac{1}{2}}\frac{\bm{\mu}_{ij}\cdot\bm{\epsilon}_{\kl}}{\hbar}$ are the light-matter coupling constants and $\bm{\mu}_{ij} =\bra{i}\bm{\hat{\mu}}\ket{j}$ are the transition dipole matrix elements, assumed real.

The equations of motion of the system-bath composite $\hat{\rho}=\hat{\rho}_S\otimes \hat{\rho}_R$ in the interaction picture are given by the Liouville Von-Neuman equation \cite{breuer_theory_2007,blum_density_2012}
\beq
\dot{\hat{\rho}}(t)=-i[\hat{V}_{SR}^I(t),\hat{\rho}(0)] -\int^t_0 dt'[\hat{V}_{SR}^I(t),[\hat{V}_{SR}^I(t'),\hat{\rho}(t')]]
\label{eq:LVNE}
\eeq
If the system and bath are initially in a separable state $\hat{\rho}(0)=\hat{\rho}_S(0)\otimes \hat{\rho}_R(0)$, the Born approximation states that, for weak system-bath coupling (which is valid for the natural light excitation of LHC's) they remain in a separable state at all times.
Furthermore, the small system produces no back reaction on the large bath.

Applying the Born approximation to \cref{eq:LVNE} and tracing over the bath coordinates gives the equations of motion for the reduced system density matrix $\hat{\rho}_S$:
\beq
\dot{\hat{\rho}}_S(t)=-i \textrm{Tr}_R[\hat{V}_{SR}^I(t),\hat{\rho}_S(0)\otimes\hat{\rho}_R(0)] -\int^t_0 dt'\textrm{Tr}_R[\hat{V}_{SR}^I(t),[\hat{V}_{SR}^I(t'),\hat{\rho}_S(t')\otimes\hat{\rho}_R(t')]]
\label{eq:LVN}
\eeq
where $\text{Tr}_R$ denotes a trace over the radiation field. While this equation is similar  to the standard master equation for stationary baths \cite{breuer_theory_2007,blum_density_2012},  the radiation field density matrix now carries  an explicit time-dependence due to the interaction with the environment.

Given Eq. (\ref{eq:VSB}), the double commutator in \cref{eq:LVN} gives products of $\hat{V}_{SR}^I$ at two different times, with a typical term  of the form:
\beq
\int^t_0 dt' \sum_{i\leq j}\sum_{l\leq m}\sum_{\kl}\sum_{\klp}g_{\kl}^{(i,j)}g_{\klp}^{(l,m)} \langle \akl \hat{a}^\dagger_{\bm{k'}\lambda'} \rangle (t')\hat{\sigma}_{i,j}\hat{\sigma}_{m,l} \hat\rho_S(t')e^{-i(\omega_{ij}-\nu_k)t'+i(\omega_{lm}-\nu_{k'})t}~.
\label{eq:DCTerm1}
\eeq
Here $\hat{\sigma}_{i,j} = \ketbra{i}{j}$ is the ``quantum jump operator'' from state $\ket{j}$ to state $\ket{i}$.

Using the commutator algebra of bosonic creation and annihilation operators, and the identity $\akl\hat{a}^\dagger_{\bm{k'}\lambda'} = (1+\hat{n}_{\kl})\delta_{\bm{k},\bm{k'}}\delta_{\lambda,\lambda'}$, the trace over the bath in \cref{eq:DCTerm1} is obtained as:
\beq
\langle \akl \hat{a}^\dagger_{\bm{k'}\lambda'} \rangle (t)=Tr_R\left\{ \akl \hat{a}^\dagger_{\bm{k'}\lambda'} \hat\rho_R(t)\right\}=\delta_{\bm{k},\bm{k'}}\delta_{\lambda,\lambda'}\left(1+n_{\bm{k}\lambda}(t)\right)
\label{eq:aTrace}
\eeq
Here we have used the normalization of the bath density matrix $\textrm{Tr}_R\{\hat\rho_R(t)\}=1$,   \cref{eq:nt}, and the linearity of the trace.
Substituting \cref{eq:aTrace} into \cref{eq:DCTerm1} gives

\beq
\int^t_0 dt' \sum_{i\leq j}\sum_{l\leq m}\sum_{\kl}g_{\kl}^{(i,j)}g_{\kl}^{(l,m)} (1+n_{\kl}(t'))\hat{\sigma}_{i,j}\hat{\sigma}_{m,l} \hat\rho_S(t')e^{-i(\omega_{ij}-\nu_k)t'+i(\omega_{lm}-\nu_k)t}
\label{eq:DCTerm2}
\eeq

Taking the continuum limit of the $\bm{k}$ summation
\beq
\sum_{\bm{k}} \to \frac{2V}{(2\pi)^3}\int^{2\pi}_0 d\phi\int^\pi_0d\theta\sin(\theta)\int^\infty_0dk \,k^2
\label{eq:kcont}
\eeq
and noting that for an isotropic and unpolarized field $n_{\kl}$ depends only on $k$ and not on the angular or polarization coordinates, \cref{eq:DCTerm2} can be rearranged to yield

\begin{multline}
\label{eq:DCTerm3}
\frac{1}{8\pi^3\epsilon_0}\int^t_0dt'\sum_{i \leq j}\sum_{l\leq m}\hat\sigma_{i,j}\hat\sigma_{m,l}\hat\rho_S(t')e^{-i(\omega_{ij}t'-\omega_{lm}t)}\\
\times\left[\sum_{\lambda}\int^{2\pi}_0d\phi\int^\pi_0d\theta \sin(\theta)(\bm\mu_{ij}\cdot\bm\epsilon_{\kl})(\bm\mu_{lm}\cdot\bm\epsilon_{\kl})\right]\left[\int^\infty_0 dk k^2\nu_k(1+n_{\kl}(t'))e^{i\nu_k(t-t')}\right]
\end{multline}

The summation over the angular and polarization coordinates in \cref{eq:DCTerm3} can be evaluated to give $\bm\mu_{ij}\cdot\bm\mu_{lm}$.
Changing variables from $k$ to $\nu_k =ck$, \cref{eq:DCTerm3} can be written in the following form:

\beq
\sum_{i\leq j}\sum_{l\leq m} \frac{\bm\mu_{ij}\cdot\bm\mu_{lm}}{8\pi^3\epsilon_0c^3}e^{i(\omega_{ij}-\omega_{lm})t}\int^\infty_0 d\nu_k\nu_k^3\int^t_0dt' e^{i(\omega_{ij}-\nu_k)(t-t')}(1+n_{\kl}(t'))\hat\sigma_{i,j}\hat\sigma_{l,m}\hat\rho_S(t')
\label{eq:DCTerm4}
\eeq
The exponential factor in the integrand oscillates rapidly at $\nu_k\ne \omega_{ij}$, so
provided that $n_{\kl}$ varies slowly near $\omega_{ik}$ at all times, we can make the Wigner-Weisskopff approximation by setting \cite{scully_quantum_1997} 
\beq
\int^\infty_0 d\nu_k\nu_k^3 e^{-i\nu_k(t-t')}  \rightarrow  \omega_{ij}^3 \int_0^\infty d\nu_k e^{-i\nu_k(t-t')} 
\label{eq:WWA}
\eeq
giving
\beq
\sum_{i\leq j}\sum_{l\leq m} \frac{\bm\mu_{ij}\cdot\bm\mu_{lm}\omega_{ij}^3}{8\pi^3\epsilon_0c^3}e^{i(\omega_{ij}-\omega_{lm})t}\int^t_0dt'(1+n_{\kl}(t')) \hat\sigma_{i,j}\hat\sigma_{m,l}\hat\rho_S(t') e^{i\omega_{ij}t'}\int^\infty_0d\nu_ke^{-i\nu_k(t-t')}
\label{eq:DCTerm5}
\eeq

The $\nu_k$ integral in \cref{eq:DCTerm5} can now be evaluated as $\pi\delta(t-t')+iP(1/(t-t'))$ where $P$ denotes the Cauchy Principal Part.
Neglecting the small Lamb shift due to the imaginary part of Eq. (\ref{eq:WWA}), and doing the  time integral gives
\beq
\sum_{i\leq j}\sum_{l \leq m}\frac{\mu_{ij}\mu_{lm}p_{ij,lm}\omega_{ij}^3}{8\pi^2\epsilon_0c^3}(1+n_{k_{ij}\lambda}(t))\hat\sigma_{ij}\hat\sigma_{m,l}\hat\rho_S(t)e^{i(\omega_{lm}-\omega_{ij})t}
\label{eq:IntTerm}
\eeq
where alignment parameters, $p_{ij,lm}$, for the transition dipole moments have been defined:

\beq
p_{ij,lm}=\frac{\bm\mu_{ij}\cdot\bm\mu_{lm}}{\mu_{ij}\mu_{lm}}
\label{eq:p}
\eeq

Transforming \cref{eq:IntTerm} back into the Schr\"odinger picture eliminates the oscillating phase factor and yields the same master equations as previously reported for stationary fields \cite{tscherbul_partial_2015}, but with  time dependent occupation numbers.  That is, following the same approach as in Ref. \cite{tscherbul_partial_2015} we arrive at the same master equations [Eq. (17) in  Ref. \cite{tscherbul_partial_2015}] with the following substitution.
\beq
r_{ij}=\gamma_{ij}\overline{n}_{\omega_{ij}} \to r_{ij}(t)=\gamma_{ij}\overline{n}_{\omega_{ij}}(t) \label{eq23}
\eeq
where $r_{ij}(t)$ is the pumping rate from level $g_i$ in the ground state manifold to state $e_j$ in the excited state manifold.

\section{V-System Master Equations}
\label{sec:Vsys}
\begin{figure}[ht!]
	\centering
	\includegraphics[width=0.45\textwidth]{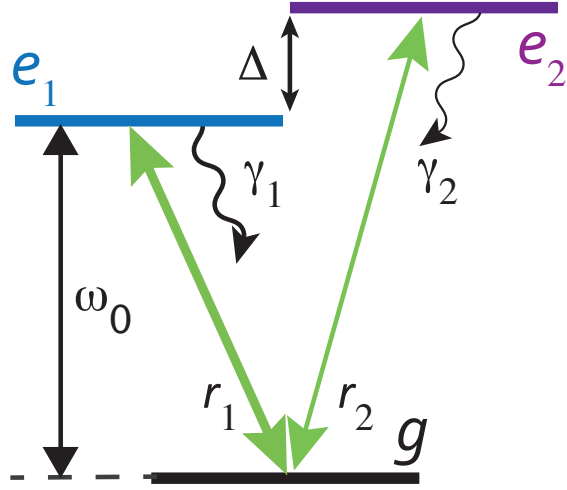}
	\renewcommand{\figurename}{Fig.}
	\caption{Schematic representation of a V-type System. $\Delta$ is the excited state splitting, $\gamma_i$ is the radiative line-width, and $r_i$ is the incoherent pumping rate of excited state $\ket{e_i}$.}
\label{fig:schem}
\end{figure}

Consider now a V-system with one  ground state  and two  excited states  (as shown in \cref{fig:schem}) coupled to a time dependent incoherent radiation field.
Since there is only one ground state in the V-system, we suppress the first index ($i$) which specifies the ground state in the pumping rate $r_{ij}(t)$ and spontaneous emission rate $\gamma_{ij}$.
That is $r_{ij}(t) \to r_j(t)$ gives the pumping rate from the ground state $g$ to the excited state $e_j$ and $\gamma_{ij}\to\gamma_j$ gives the spontaneous decay rate from excited state $e_j$ to the ground state $g$.
The master equation for such a system is given by
\begin{subequations}
\beq
\dot{\rho}_{e_j e_j} =-(r_j(t) + \gamma_j)\rho_{e_j e_j} +r_j (t)\rho_{gg}-p(\sqrt{r_1(t)r_2(t)}+\sqrt{\gamma_1\gamma_2})\rho_{e_1e_2}^R
\label{eq:PQME}%
\eeq
\beq
\begin{split}
\dot{\rho}_{e_1e_2}= &-\frac{1}{2}(r_1(t)+r_2(t)+\gamma_1+\gamma_2)\rho_{e_1e_2}-i\rho_{e_1 e_2}\Delta
                \\&+\frac{p}{2}\sqrt{r_1(t) r_2(t)}(2\rho_{gg}-\rho_{e_1
                  e_1}-\rho_{e_2
                  e_2}) -\frac{p}{2}\sqrt{\gamma_1\gamma_2}(\rho_{e_1
                  e_1}+\rho_{e_2 e_2})
\label{eq:CQME}%
\end{split}
\eeq
\label{eq:QME}%
\end{subequations}
where $\rho_{e_1e_2}^R$ is the real part of the off-diagonal (coherence) density matrix element between levels $e_1$ and $e_2$.
In \cref{eq:QME}, spontaneous emission processes are governed by the radiative decay widths of the excited states, $\gamma_i=\omega_{g e_i}^3|\mu_{g e_i}|^2/(3\pi\epsilon_0c^3)$,
$\Delta=\omega_{e_1 e_2}$ gives the excited state splitting and $p=\bm{\mu}_{g e_1}\cdot\bm{\mu}_{g e_2}/(|\mu_{g e_1}| |\mu_{g e_2}|)$ measures the alignment of the $\ket{g} \leftrightarrow \ket{e_i}$ transition dipole moments, $\bm{\mu}_{g e_i}$.
Absorption and stimulated emission processes are parametrized by time-dependent incoherent pumping rates of the $\ket{g} \leftrightarrow \ket{e_i}$ transitions, $r_i(t)=\gamma_i \nbar(t)$. 
Here we neglect  environment-induced dephasing and relaxation processes, assuming that the rates of excited state relaxation and dephasing are small compared to those of the radiative processes (absorption, decay and stimulated emission) \cite{dodin_quantum_2016}.
These effects can, however be included by generalizing this approach \cite{tscherbul_long-lived_2014}.

As in the case of stationary pumping rates \cite{dodin_quantum_2016}, applying the conservation of population constraint, $\rho_{gg}=1-\rho_{e_1e_1}-\rho_{e_2e_2}$, and transforming into the Liouville space representation with state vector $\bm{x}=[\rho_{e_1e_1}, \rho_{e_2e_2}, \rho_{e_1e_2}^R,\rho_{e_1e_2}^I]^T$ yields \cref{eq:QME} in a vector form:
\begin{subequations}
\beq
\frac{d}{dt} \bm{x} = A(t)\bm{x}+ \bm{d}(t)
\label{eq:MLRME}%
\eeq
\beq
A(t)= \left(
\begin{array}{cccc}
-(2r_1(t)+\gamma_1) & -r_1(t) & -p\sqrt{\gamma_1\gamma_2}\left(1+\nbar(t)\right) & 0 \\
-r_2(t) & -(2r_2(t)+\gamma_2) & -p\sqrt{\gamma_1\gamma_2}\left(1+\nbar(t)\right) & 0 \\
-\frac{p\sqrt{\gamma_1\gamma_2}}{2}\left(1+3\nbar(t)\right) & -\frac{p\sqrt{\gamma_1\gamma_2}}{2}\left(1+3\nbar(t)\right) & -\gbar(1+\nbar(t)) & \Delta \\
0 & 0 & -\Delta & -\gbar(1+\nbar(t))
\end{array}\right)
\label{eq:Amat}%
\eeq
\beq
\mathbf{d}(t)=\left(\begin{array}{c}
r_1(t) \\ r_2(t) \\ p\sqrt{r_1(t)r_2(t)}\\0
\end{array} \right)
\label{eq:dvec}%
\eeq
\label{eq:VME}%
\end{subequations}
where $\gbar=\half(\gamma_1+\gamma_2)$ is the arithmetic mean of the spontaneous decay rates of the excited state manifold, and we have used the fact that $r_i(t)/\gamma_i=\nbar(t)$.

Rewriting \cref{eq:dvec} in the form $\mathbf{d}(t)=[\gamma_1, \gamma_2, p\sqrt{\gamma_1\gamma_2},0]^T \nbar(t)$ shows that the time-varying field pumps the system to the same statistical mixture, $\rho_d$, in the excited manifold as does the stationary field \cite{dodin_quantum_2016}, where
\beq
\rho_d \propto (1-p)(\gamma_1\ket{e_1}\bra{e_1} + \gamma_2\ket{e_2}\bra{e_2})+p\ket{\phi_+}\bra{\phi_+}.
\label{eq:rhod}
\eeq
Here $\ket{\phi_+}=(1/\sqrt{2\gbar})(\sqrt{\gamma_1}\ket{e_1}+\sqrt{\gamma_2}\ket{e_2})$ is an in-phase coherent superposition of excited energy eigenstates.
However, in contrast to the stationary field case, the rate of excitation into this statistical mixture varies with time. 
Hence, although these equations look similar to the stationary case, the time-varying excitation rate in this case  produces very different dynamics than one sees in the stationary rate case.

Provided the weak-pumping limit ($\nbar(t) \ll 1$) is satisfied at all times, the coefficient matrix $A(t)$ \cref{eq:Amat} can be perturbatively expanded in $\nbar(t)$.
This yields a time-independent coefficient matrix $A^{(0)}$ to zeroth order in $\nbar(t)$, allowing the time-dependent contributions to $A(t)$ to be treated through a perturbative expansion:

\begin{subequations}
\beq
A(t)=A^{(0)}+\nbar(t) A^{(1)}
\label{eq:Asum}
\eeq
\beq
A^{(0)}= \left(
\begin{array}{cccc}
-\gamma_1 & 0 & -p\sqrt{\gamma_1\gamma_2}& 0 \\
0 & -\gamma_2 & -p\sqrt{\gamma_1\gamma_2} & 0 \\
-\frac{p\sqrt{\gamma_1\gamma_2}}{2} & -\frac{p\sqrt{\gamma_1\gamma_2}}{2} & -\gbar & \Delta \\
0 & 0 & -\Delta & -\gbar
\end{array}\right)
\label{eq:A0}
\eeq
\label{eq:Apert}
\end{subequations}

As a result, this yields  a linear system of constant coefficient ordinary differential equations with a time dependent driving term.
Applying the initial conditions appropriate to excitation from a molecule in the ground state, $\rho_{gg}(0)=1$ or $\bm{x_0}=\bm{0}$, the dynamics of the V-system is given by the variation of parameters solution \cite{boyce_elementary_2009}
\beq
\mathbf{x} = \int^t_0 ds e^{A^{(0)}(t-s)}\mathbf{d}(s) \to \sum_{i=1}^4 \int^t_0 ds (\mathbf{v}_i\cdot\mathbf{d}(s))e^{\lambda_i (t-s)} \mathbf{v}_i
\label{eq:GenSol}
\eeq
where $\lambda_i$ is the $i^{th}$ eigenvalue of $A^{(0)}$ with corresponding eigenvector $\mathbf{v_i}$.
Eq. \ref{eq:GenSol} relates the eigenvalues, $\{\lambda_i\}$, of $A^{(0)}$ to the timescales of the system's evolution $\tau_i= \Real(\lambda_i)^{-1}$ and to the frequencies of its oscillations $\omega_i =\Imag(\lambda_i)$. 
This solution is similar to that obtained in \cite{dodin_quantum_2016} with the crucial modification that $\mathbf{d}(s)$ is explicitly time dependent. 

Since $A^{(0)}$ is time independent, its eigenvalues, eigenvectors  and normal modes are identical to those calculated in the stationary field case \cite{dodin_quantum_2016}, with
\beqs
\beq
\lambda_i= -\gbar \pm \Delta_p \sqrt{\zeta^2-1} \sqrt{\frac{1 \pm \sqrt{1+\eta^2}}{2}}
\label{eq:lamDO}%
\eeq
\beq
\zeta =\frac{\gbar}{\Delta_p}
\label{eq:zeta}%
\eeq
\beq
\eta =\frac{\Delta|\gamma_1-\gamma_2|}{|\gbar^2-\Delta_p^2|}
\label{eq:eta}%
\eeq
\eeqs
where $\Delta_p =\sqrt{\Delta^2+(1-p^2)\gamma_1\gamma_2}$.

The overdamped ($\Delta_p/\bar\gamma \ll 1$) and underdamped ($\Delta_p/\bar\gamma \gg1$) regimes of a V-system excited by a field with a finite turn on time $\tau_r$ of the form \cref{eq:nt} are discussed below.
Since $\Delta_p \geq \Delta$ where the latter is the excited state level splitting, the overdamped region would be relevant to, e.g., large molecules whereas the underdamped region would correspond, e.g., to small molecules.

\section{Overdamped Regime $\frac{\Delta_p}{\gbar} \ll 1$}
\label{sec:OD}
In the overdamped regime, where  $\zeta=\frac{\gbar}{\Delta_p} \gg 1$ of \cref{eq:lamDO}, the eigenvalues take the simplified form \cite{dodin_quantum_2016}
\beqs
\beq
\label{eq:Olam1}
\lambda_1=-2\gbar
\eeq
\beq
\lambda_2=-\frac{\Delta_p^2}{2\gbar}
\label{eq:Olam2}
\eeq
\beq
\lambda_{3,4}=-\gbar~,
\label{eq:Olam34}
\eeq
\label{eq:Olams}
\eeqs
while the normal modes are given by \cite{dodin_quantum_2016}
\beqs
\beq
\mathbf{v}_1 \propto [r_1,r_2,p\sqrt{r_1r_2},0]
\label{eq:Ov1}%
\eeq
\beq
\mathbf{v}_2 \propto [r_2,r_1,-p\sqrt{r_1r_2},0]
\label{eq:Ov2}%
\eeq
\beq
\mathbf{v}_3 \propto [0,0,0,1]
\label{eq:Ov3}%
\eeq
\beq
\mathbf{v}_4 \propto [1,-1,-\frac{\gamma_1-\gamma_2}{p\sqrt{\gamma_1\gamma_2}},0]~.
\label{eq:Ov4}%
\eeq
\label{eq:Ovs}%
\eeqs

Substituting \cref{eq:Ovs,eq:Olams} for the eigenvalues and normal modes,  and \cref{eq:nt} for the occupation number, into the variation of parameters solution \cref{eq:GenSol} and doing the  exponential integrals yields the V-system dynamics
\beqs
\beq
\begin{split}
\rho_{e_ie_i}(t)
= \frac{1}{2\gbar} &\bigg\{\frac{r_j}{\frac{\Delta_p^2}{2\gbar}-\alpha}\left[\frac{\Delta_p^2}{2\gbar}\left(1-e^{-\alpha t}\right) -\alpha\left(1-e^{-\frac{\Delta_p^2}{2\gbar}t}\right)\right]\\
&+\frac{r_i}{2\gbar-\alpha}\left[2\gbar\left(1-e^{-\alpha t}\right) -\alpha\left(1-e^{-2\gbar t}\right)\right]\bigg\}
\end{split}
\label{eq:ODP}%
\eeq
\beq
\begin{split}
\rho_{e_1e_2}(t)=
\frac{p\sqrt{r_1r_2}}{2\gbar} &\bigg\{\frac{1}{\frac{\Delta_p^2}{2\gbar}-\alpha}\left[\frac{\Delta_p^2}{2\gbar}\left(1-e^{-\alpha t}\right) -\alpha\left(1-e^{-\frac{\Delta_p^2}{2\gbar}t}\right)\right]\\
&-\frac{1}{2\gbar-\alpha}\left[2\gbar\left(1-e^{-\alpha t}\right) -\alpha\left(1-e^{-2\gbar t}\right)\right]\bigg\}
\end{split}
\label{eq:ODC}%
\eeq
\label{eq:ODamp}%
\eeqs
where $i,j=1,2$ and $i\neq j$ and $r_i=\lim_{t \to \infty}r_i(t)=\gamma_i\nbar$ is the steady state incoherent pumping rate. Here $\alpha$, we recall from \cref{eq:ft}, defines the turn on time $\tau_r =1/\alpha$.

\begin{figure}[ht]
	\centering
	\includegraphics[width=\textwidth, trim = 0 0 0 0]{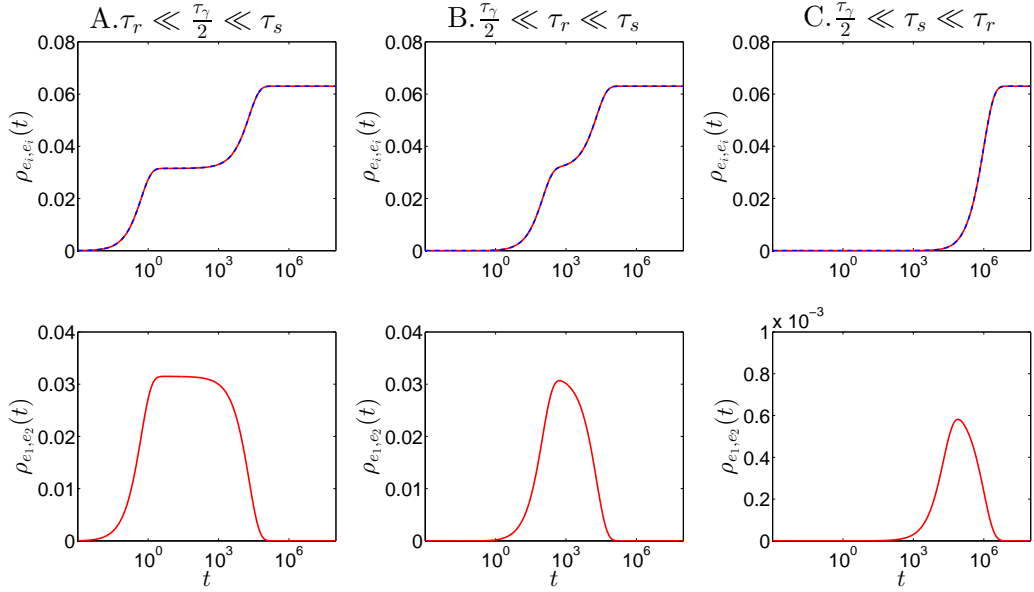}
	\renewcommand{\figurename}{Fig.}
	\caption{Evolution of populations and coherences of an overdamped V-system ($\frac{\Delta_p}{\gbar} \ll 1$) evaluated with aligned transition dipole moments ($p=1$). Here $\gamma_1=1.0=\gamma_2$ and $\Delta=0.001$. Three different turn on regimes are shown. Panels A  show the ultrafast turn on of the field with $\tau_r=\times10^{-3}\tau_\gamma$ while Panels B and C show the intermediate ($\tau_r=100\tau_\gamma=5\times10^{-5}\tau_s$) and slow ($\tau_r= 20\tau_s$) turn on regimes, respectively. Note the difference in y-axis scales for the different coherence plots.}
\label{fig:1}
\end{figure}

\Cref{eq:ODP,eq:ODC} show that the steady state behavior is given by $\lim_{t\to\infty}\bm{x}(t)=[\nbar,\nbar,0,0]$, is independent of $\alpha \neq0$, and is identical to the steady state obtained for stationary fields \cite{dodin_quantum_2016}.
However, the non-equilibrium behavior of the system and the maximal coherence can be markedly different for time-varying fields, as shown in \cref{fig:1}.

To illustrate this consider a system where the radiation field turn on time $\tau_r=1/\alpha$ differs significantly from $\tau_\gamma/2=1/2\gbar$ and $\tau_s=2\gbar/\Delta_p^2$.
Since the overdamped regime imposes the inequality $\tau_s \gg \tau_\gamma/2$, this corresponds to a separation of timescales $\{\tau_\gamma/2,\tau_s,\tau_r\}$.
There are three possible time-orderings: the sudden turn on ($\tau_r \ll\tau_\gamma/2\ll\tau_s$), the slow turn on ($\tau_\gamma/2\ll\tau_s\ll\tau_r$) and the intermediate ($\tau_\gamma/2\ll\tau_r\ll\tau_s$) regimes.

For the sudden turn on of the field, \cref{eq:ODamp} simplifies to
\beqs
\beq
\rho_{e_ie_i}(t)=\frac{1}{2\gbar}\left[ r_i\left(1-e^{-2\gbar t}\right)+r_j\left(1-e^{-\frac{\Delta_p^2}{2\gbar}t}\right) \right]
\label{eq:ODPSt}%
\eeq
\beq
\rho_{e_1e_2}(t)=\frac{p\sqrt{r_1r_2}}{2\gbar}\left(e^{-\frac{\Delta_p^2}{2\gbar}t}-e^{-2\gbar t}\right)
\label{eq:ODCSt}%
\eeq
\label{eq:ODSt}%
\eeqs
under a binomial expansion to lowest contributing order in ${\tau_r}/{\tau_{s}}\ll1$.
\Cref{eq:ODSt} is independent of $\alpha$ (and hence $\tau_r$) and is identical to the expression derived for stationary fields, the $\tau_r \to 0$ limit \cite{dodin_quantum_2016}.
The large quasistationary coherences characteristic of this regime can clearly be seen in subplot A of \cref{fig:1}.
That is, if the field is turned on faster than the fastest characteristic time scale of the system $\tau_\gamma/2 = 1/2\gbar$ then the dynamics of the system are well approximated by the stationary field solution, \cref{eq:ODSt}.
In particular, the coherences approach the same maximal value of $\frac{p\sqrt{r_1r_2}}{2\gbar}$ in the interval $\tau_{\gamma}/2 < t <\tau_s=\frac{2\gbar}{\Delta_p^2}$ as in the stationary field approximation.
In this limit, the field reaches its steady state much faster than the system evolves, and so the very short-lived transient behavior of the field is not reflected in the evolution of the system.
Intuitively, in this regime the V-system does not evolve under the transient field. Instead, it evolves only under the steady state field.

By contrast, if the field is turned on very slowly, $\tau_\gamma/2\ll\tau_s\ll\tau_r$, the stationary field solution is a very poor approximation for the system dynamics.
Taking a binomial expansion to lowest contributing order in ${\tau_{s}}/\tau_r \ll 1$, \cref{eq:ODamp} can be rewritten as
\beqs
\beq
\rho_{e_ie_i}(t)=\nbar\left(1-e^{-\alpha t}\right)
\label{eq:ODPSl}%
\eeq
\beq
\rho_{e_1e_2}(t)=\frac{p\alpha\sqrt{r_1r_2}}{\Delta_p^2}\left(e^{-\frac{\Delta_p^2}{2\gbar}t}-e^{-\alpha t}\right)
\label{eq:ODCSl}%
\eeq
\label{eq:ODSl}%
\eeqs
The dependence on the incoherent pumping rates, $r_i$, in \cref{eq:ODSl} is contained in the mean thermal occupation of the field $\nbar =r_i/\gamma_i$.
To appreciate this result, note that when the field is turned on adiabatically, the dynamics of the system closely resemble the incoherent excitation produced by Pauli rate law dynamics. 
The rate law predicts populations evolving to an equilibrium value of $\nbar$ as $\rho_{e_ie_i}=\nbar(1-e^{-\gamma_it})$.
This is  similar to \cref{eq:ODPSl}, where the population of the excited states equilibrates to the same value $\nbar$ at the rate $\alpha$.
\Cref{eq:ODPSl} may also be rewritten as $\rho_{e_ie_i}(t)=\nbar(t)$ by substituting \cref{eq:nt}, indicating that the system is in equilibrium at all times under the slowly-varying field.
Furthermore, \cref{eq:ODCSl} shows a suppression of the coherences by a factor of $\frac{\tau_s}{\tau_r}\ll1$ in comparison to excitation by a field with a very fast turn on time.
Alternatively, writing this in terms of the characteristic timescales of the system
\beq
\max\{|\rho_{e_1e_2}^{Slow}(t)|\}=\frac{\tau_s}{\tau_r}\max\{|\rho_{e_1e_2}^{Fast}|\}
\label{eq:MaxComp}
\eeq
where $\frac{\tau_s}{\tau_r}=\frac{\alpha}{\lambda_2} \ll1$.
The difference in coherence amplitude between the fast and slow turn on of the radiation field can be seen in comparing subplots A and C of \cref{fig:1}.

The suppression of the coherence amplitude under adiabatic turn-on of the field can be understood by considering the evolution of individual components of $\rho_d$ \cref{eq:rhod} and the interactions between them.
The coherences originate from the in-phase $\ket{\phi_+} = (1/\sqrt{2\gbar})(\sqrt{\gamma_1}\ket{e_1}+\sqrt{\gamma_2}\ket{e_2})$ superposition prepared by the incident field.
This superposition  collapses to  an equally populated incoherent mixture of  excited states $\rho_{eq}=\nbar(\ket{e_1}\bra{e_1}+\ket{e_2}\bra{e_2})$ over  a time-scale  $\tau_s$ \cite{tscherbul_long-lived_2014,dodin_quantum_2016}.
Furthermore, the population of excited states enhances the decay of in-phase superpositions through the increased rate of decay processes.
This disproportionately affects the in-phase superpositions since they exhibit constructive interference in the decay processes which is reflected in the terms proportional to $\rho_{e_ie_i}$ in the coherence master equations \cref{eq:QME}.
Since the $\ket{\phi_+}$ superpositions generated by the incoherent field decay at a time-scale $\tau_s$ leaving behind an incoherent mixture of excited states, the $\ket{\phi_+}$ states prepared at later times will decay faster than those prepared at earlier times due to the increased population of the excited states.
As a result, appreciable amounts of $\ket{\phi_+}$ never accumulate in the system, leading to a heavy suppression of the coherences.
This also accounts for the decay of the coherences with time-scale $\tau_r$ [\cref{eq:ODCSl}].
The population of excited states on this time-scale lead to an increase in the decay rate of the coherent superpositions on a time-scale $\tau_r$. 
This ultimately leads to the decay of the $\ket{\phi_+}$ components on the radiation field turn on time.

Hence,, when the system is excited by a field that is turned on very slowly compared to the system's longest time scale (here $\tau_s=2\gbar/\Delta_p^2$), it will evolve in constant equilibrium with the field, producing the incoherent instantaneous steady state $\bm{x}(t)=[\nbar(t),\nbar(t),0,0]$ at all times.

For completeness, consider a field in the intermediate turn on regime, $\tau_\gamma/2 \ll \tau_r \ll \tau_s$.
Proceeding through a binomial expansion, as in the earlier cases, \cref{eq:ODamp} reduces to
\beqs
\beq
\rho_{e_ie_i}(t)=\frac{1}{2\gbar}\left[ r_i\left(1-e^{-2\alpha t}\right)+r_j\left(1-e^{-\frac{\Delta_p^2}{2\gbar}t}\right) \right]
\label{eq:ODPInt}%
\eeq
\beq
\rho_{e_1e_2}(t)=\frac{p\sqrt{r_1r_2}}{2\gbar}\left(e^{-\frac{\Delta_p^2}{2\gbar}t}-e^{-\alpha t}\right)
\label{eq:ODCInt}%
\eeq
\label{eq:ODInt}%
\eeqs
This implies that, in the intermediate regime the system displays the same maximal coherence as in the fast turn on regime.
However, the time scale over which  it approaches its quasistationary state becomes $\tau_r=1/\alpha$ rather than $\tau_\gamma/2={1}/{2\gbar}$.
When the turn on time is slower than the decay time ($\tau_s$) of $\ket{\phi_+}$, the radiation field reaches a steady state before the $\ket{\phi_+}$ excitations decay appreciably.
In contrast to the adiabatic turn on case, the survival of the early coherences in the intermediate regime allows for the maintenance of coherences from excitations at later times.
This leads to the same maximal coherence in the  intermediate regime as in the sudden turn-on case.
However, note that although most of the coherences are generated in the timescale $\tau_r$ they decay at the same time as those generated for the sudden turn-on, as $\tau_s$ rather than ($\tau_r+\tau_s$) as may be expected a priori.
This indicates that excitations to $\ket{\phi_+}$ generated at later times have a shorter decay time than those generated at earlier times.
This occurs through the same mechanism as the decay of coherences on a time scale of $\tau_r$ in the adiabatic turn-on regime.
When the $\ket{\phi_+}$ excitations generated at earlier times decay to incoherent mixtures of excited states at time $\tau_s$ the increase in excited state population leads to an increase in the decay rate of the coherences.
This leads to a ``cascade'' in which the rate of decay of the coherences increases as more in-phase superpositions decay to incoherent mixtures of the excited eigenstates.

\section{Underdamped Regime $\frac{\gbar}{\Delta_p} \ll 1$}
\label{sec:UD}
A V-system in the underdamped regime is characterized by a very small damping coefficient, $\zeta=\frac{\gbar}{\Delta_p} \ll1$.
Taking the corresponding limit of \cref{eq:lamDO} gives the eigenvalues of an underdamped V-system \cite{dodin_quantum_2016}:

\beqs
\beq
\lambda_1 = -\gamma_1
\label{eq:lam1}
\eeq
\beq
\lambda_2 = -\gamma_2
\label{eq:lam2}
\eeq
\beq
\lambda_{3,4} = -\gbar \pm i\Delta_p
\label{eq:lam34}
\eeq
\label{eq:lams}
\eeqs

Substituting \cref{eq:lams} into \cref{eq:Amat}, one  finds the corresponding normal modes \cite{dodin_quantum_2016}

\beqs
\beq
\mathbf{v_1} \propto [1,0,0,\frac{p\sqrt{\gamma_1\gamma_2}}{2\Delta_p}]
\label{eq:v1}
\eeq
\beq
\mathbf{v_2} \propto [0,1,0,\frac{p\sqrt{\gamma_1\gamma_2}}{2\Delta_p}]
\label{eq:v2}
\eeq
\beq
\mathbf{v_3} \propto [0,0,1,1]
\label{eq:v3}
\eeq
\beq
\mathbf{v_4} \propto [0,0,1,-1]
\label{eq:v4}
\eeq
\label{eq:vs}
\eeqs

The general solution,  obtained using \cref{eq:GenSol}, is
\beqs
\beq
\rho_{e_ie_i}(t)=\frac{\nbar}{\alpha-\gamma_i}\left[\alpha(1-e^{-\gamma_i t})-\gamma_i(1-e^{-\alpha t})\right]
\label{eq:UDP}%
\eeq
\beq
\begin{split}
\rho_{e_1e_2}^R & =p\sqrt{r_1r_2}\bigg[ \frac{e^{-\gbar t}(\gbar(1-\cos(\Delta_p t))+\Delta_p\sin(\Delta_p t)}{\Delta_p^2+\gbar^2}
\\&-\frac{e^{-\gbar t}[(\alpha-\gbar)\cos(\Delta_p t)+\gbar+\Delta_p\sin(\Delta_p t)]-\alpha e^{-\alpha t}}{(\alpha-\gbar)^2+\Delta_p^2} \bigg]
\label{eq:UDCR}%
\end{split}
\eeq
\beq
\begin{split}
\rho_{e_1e_2}^I(t) &= -p\sqrt{r_1r_2}\bigg[ \frac{e^{-\gbar t}[(\Delta_p(1-\cos(\Delta_p t))-\gbar\sin(\Delta_p t)]}{\Delta_p^2+\gbar^2}\\
&- \frac{e^{-\gbar t}[\Delta_p(1-\cos(\Delta_p t))+(\alpha-\gbar)\sin(\Delta_p t)]}{\Delta_p^2+(\alpha-\gbar)^2}\\
&+\frac{1}{2(\alpha-\gamma_1)}\left[\alpha(1-e^{-\gamma_1 t})-\gamma_1(1-e^{-\alpha t})\right]\\
&-\frac{1}{2(\alpha-\gamma_2)}\left[\alpha(1-e^{-\gamma_2 t})-\gamma_2(1-e^{-\alpha t})\right]\bigg]
\label{eq:UDCI}%
\end{split}
\eeq
\label{eq:UDamp}%
\eeqs
where $\rho_{e_1e_2}^R$ and $\rho_{e_1e_2}^I$ are the real and imaginary parts of the coherence term, respectively, and where $r_i =\lim_{t\to\infty}r_i(t)=\gamma_i\nbar$ (as in the overdamped regime).

\begin{figure}[ht]
	\centering
	\includegraphics[width=\textwidth, trim = 0 0 0 0]{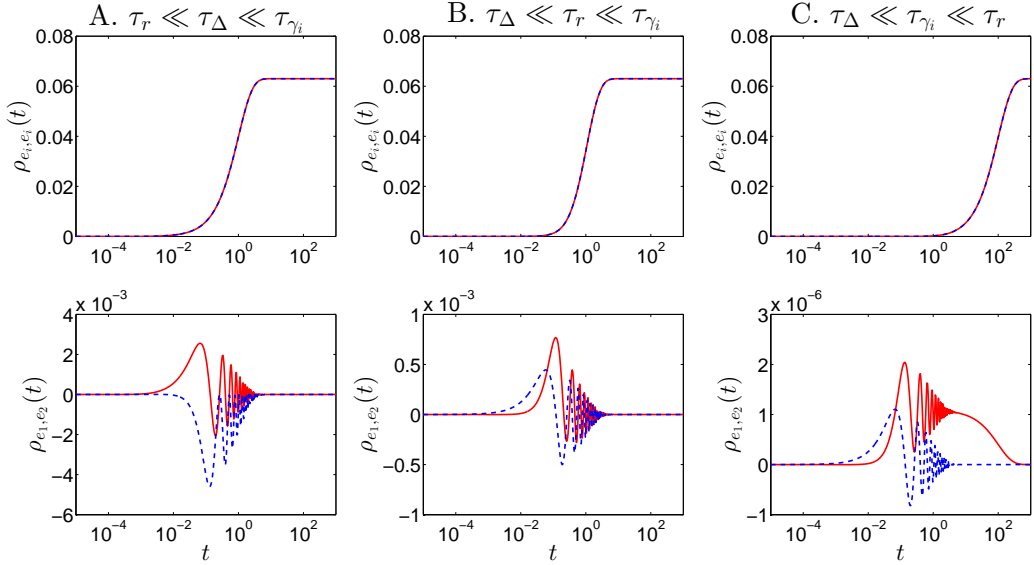}
	\renewcommand{\figurename}{Fig.}
	\caption{Evolution of populations and coherences of an underdamped V-system ($\frac{\Delta_p}{\gbar} \gg 1$) evaluated with aligned transition dipole moments ($p=1$). Here $\gamma_1=1.0=\gamma_2=\gamma$ and $\Delta=24.0$. Three different turn on regimes are shown here. Panels A show the ultrafast turn on of the field with $\tau_r=0.024\tau_\Delta$ while Panels B and C show the intermediate ($\alpha = 24\tau_\Delta$) and slow ($\alpha =100\tau_\gamma$) turn on regimes respectively. Note the difference in y-axis scales for the coherence plots. Solid red lines indicate the real part of the coherence $\rho_{e_1e_2}^R$ with the imaginary part $\rho_{e_1e_2}^I$ indicated by the dashed blue line.}
\label{fig:2}
\end{figure}

\Cref{eq:UDamp} is cumbersome and does not provide much insight into the dynamics of the system.
However, we note that the steady state of the system can easily be determined to be the incoherent mixture $\lim_{t\to\infty}\bm{x}(t)=[\nbar,\nbar,0,0]$.
This agrees with the results from both the overdamped regime and the stationary field case \cite{dodin_quantum_2016}.
In order to obtain more insight into the dynamics of the V-system we consider several cases for the turn on time.

First, consider a turn on time, $\tau_r=\alpha^{-1}$, that is  faster than all three of the system timescales, $\tau_{\gamma_i}=1/\gamma_i$ and the period of coherence oscillations $\tau_\Delta=1/\Delta_p$,
i.e.,  the fast turn on regime characterized by, $\tau_r \ll \tau_\Delta \ll \tau_{\gamma_i}$.
A binomial expansion of \cref{eq:UDamp} yields the dynamics induced by a bath with a fast turn on as:
\beqs
\beq
\rho_{e_i,e_i}(t)=\nbar (1-e^{-\gamma_it})
\label{eq:UDPSt}
\eeq
\beq
\rho_{e_1,e_2}^R(t)=\frac{p \sqrt{r_1r_2}}{\Delta_p}e^{-\gbar t}\sin(\Delta_p t)
\label{eq:UDCRSt}
\eeq
\beq
\rho_{e_1,e_2}^I(t)=\frac{p \sqrt{r_1r_2}}{\Delta_p}\left( e^{-\gbar t}(\cos(\Delta_p t)-1)-\frac{e^{-\gamma_1 t}-e^{-\gamma_2t}}{2} \right)
\label{eq:UDCISt}
\eeq
\label{eq:UDampSt}
\eeqs
As expected, this is identical to the solution derived for stationary fields \cite{dodin_quantum_2016}.
That is, if the field is turned on much faster than the characteristic timescales of the system, the stationary field solution closely approximates the evolution of the system since the field reaches its stationary state faster than the system can evolve under the transient field.

In contrast, consider a field that turns on much slower than the period of coherence oscillations, $\tau_r \gg \tau_\Delta$.
In the $\tau_\Delta \ll \tau_r$ limit \cref{eq:UDCR} for the real part of the coherence term takes on the much simpler form:
\beq
\rho_{e_1e_2}^R(t)=\frac{p\sqrt{r_1r_2}}{\Delta_p}\frac{\alpha}{\Delta_p}(e^{-\gbar t}\cos(\Delta_p t)-e^{-\alpha t})
\label{eq:UDCRSl}
\eeq
which does not depend on the value of $\tau_r$ relative to $\tau_{\gamma_i}$.
\Cref{eq:UDP,eq:UDCI} for the populations and the imaginary part of the coherence term depend on the magnitude of $\tau_r$ relative to each of the $\tau_{\gamma_i}$'s.
\Cref{eq:UDPSt} remains an accurate solution for $\rho_{e_ie_i}(t)$ provided that $\tau_{\gamma_i} \gg\tau_r$.
In the adiabatic ($\tau_{\gamma_i} \ll \tau_r$)  limit, the populations can be expressed as
\beq
\rho_{e_ie_i}(t)=\nbar(1-e^{-\alpha t})=\nbar (t)
\label{eq:UDPSl}
\eeq
\Cref{eq:UDCISt} remains a good approximation for the imaginary coherences provided that $\tau_r \ll \tau_{\gamma_i}$.
More generally, $\rho_{e_1e_2}^I(t)$ depends on the magnitude  of $\tau_r$ relative to both $\tau_{\gamma_i}$'s.
Without loss of generality, let $\gamma_1 >\gamma_2$.
This gives the dynamics of the imaginary coherences in the following cases:

\beq
\scalebox{0.97}{$\rho_{e_1e_2}(t)=\begin{cases}
-\frac{p \sqrt{r_1r_2}}{\Delta_p}\left(\frac{\alpha}{\Delta_p} e^{-\gbar t}\sin(\Delta_p t)-\frac{e^{-\gamma_1 t}-e^{-\alpha t}}{2} \right) & \mbox{if }\gamma_2 \gg \alpha \gg\gamma_1 \\
-\frac{p \sqrt{r_1r_2}}{\Delta_p}\left[\frac{\alpha}{\Delta_p} e^{-\gbar t}\sin(\Delta_p t) +\frac{\alpha}{2\gamma_1}(e^{-\gamma_1 t}-e^{-\alpha t})- \frac{\alpha}{2\gamma_2}(e^{-\gamma_2 t}-e^{-\alpha t})\right] & \mbox{if } \gamma_i \gg \alpha
\end{cases}$}
\label{eq:UDCISl}
\eeq

Significantly, in the adiabatic limit, when the turn on time of the field far exceeds the characteristic timescales of the system, the coherences in \cref{eq:UDCISl,eq:UDCRSl} are heavily suppressed by the factor of $\frac{\tau_\Delta}{\tau_r} \ll 1$ relative to the fast turn on case \cref{eq:UDCRSt,eq:UDCISt}.
Therefore, the V-system is in equilibrium with the field at all times in the adiabatic limit, producing an incoherent mixture of excited states at all times.

\begin{figure}[ht]
	\centering
	\includegraphics[width=\textwidth, trim = 0 0 0 0]{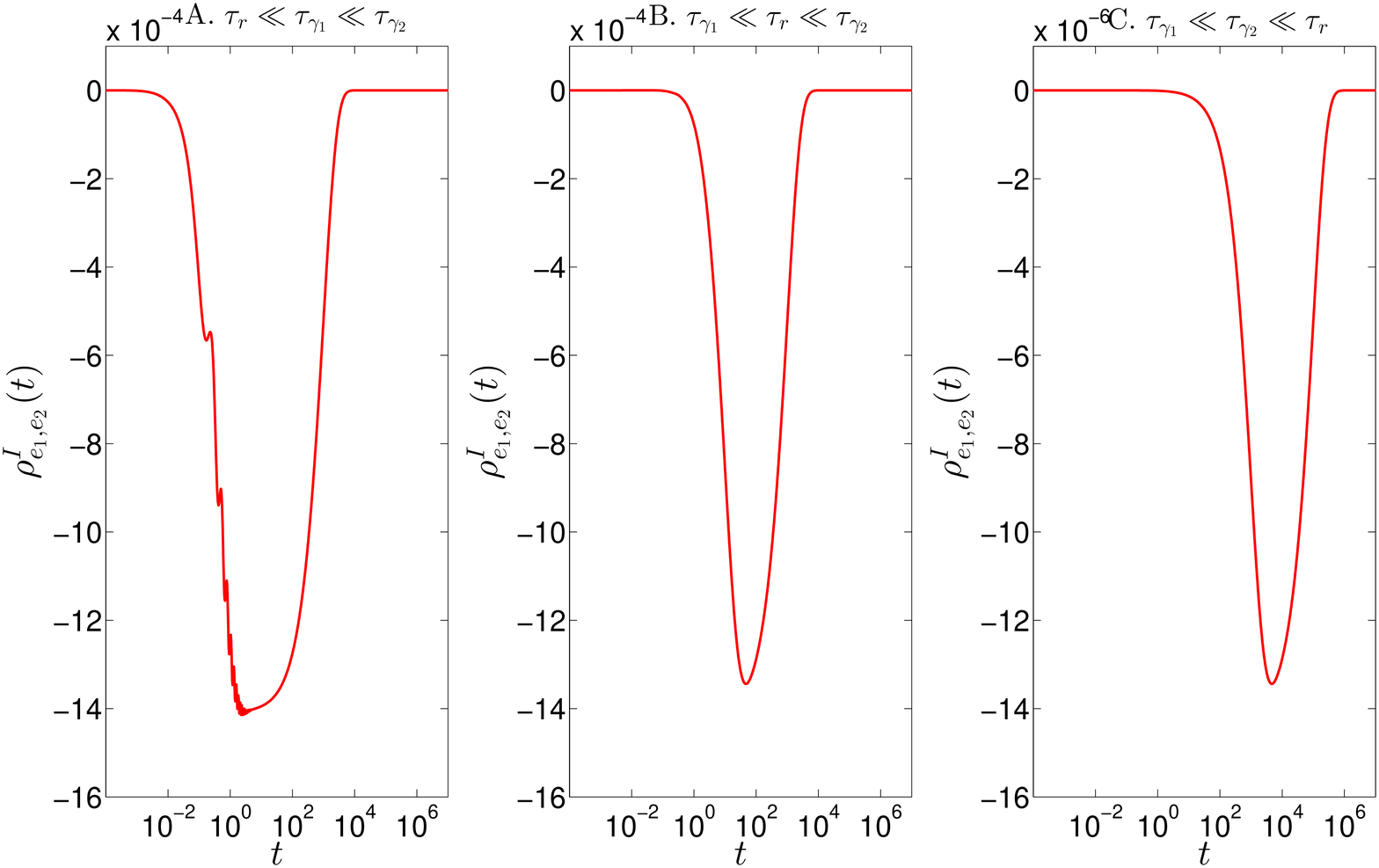}
	\renewcommand{\figurename}{Fig.}
	\caption{Evolution of the imaginary coherences of an underdamped V-system ($\frac{\Delta_p}{\gbar} \gg 1$) evaluated with aligned transition dipole moments ($p=1$). Here $\gamma_1=2.0$, $\gamma_2=10^{-3}$ and $\Delta=24.0$. Three different turn on regimes are shown here.  Panels  A shows the ultrafast turn on of the field with $\tau_r=2\times10^{-3}\tau_{\gamma_1}$ while Panels B and C show the intermediate ($\tau_r = 20\tau_{\gamma_1}=0.01\tau_{\gamma_2}$) and slow ($\tau_r = 100\tau_{\gamma_2}$) turn on regimes respectively. Note the difference in y-axis scales for the coherence plots.}
\label{fig:3}
\end{figure}

The rich dynamics of the imaginary coherences, $\rho_{e_1e_2}^I$, are hidden in \cref{fig:2} due to the assumption of equal decay rates.
\Cref{fig:3} displays the interplay between the oscillatory and quasistationary contributions more clearly by selecting highly asymmetric decay widths  $\gamma_1 \gg \gamma_2$.

Furthermore, \cref{eq:ODCSl,eq:UDCRSl,eq:UDCISl} all display the same inverse scaling of coherence amplitude with turn on time of the field.
Hence, if a radiation field is turned on slowly, the magnitude of the coherences scales inversely with the turn on time as

\beq
\max\{|\rho_{e_1e_2}|(t)\} \propto \alpha = \frac{1}{\tau_r}
\label{eq:Scale}
\eeq

This suggests that the strong coherences observed thus far \cite{tscherbul_long-lived_2014,dodin_quantum_2016} for $\Delta\neq 0$ arise due to the instantaneous turn on of the radiation field and will not be retained when considering a field with a turn on time that is slower than the radiative lifetimes of the excited states ($\tau_r \gg \tau_{\gamma_i}=1/\gamma_i$).

In summary, in both the overdamped and underdamped regions,
\textit{the Fano coherences previously computed in the study of V-systems with suddenly turned-on radiation \cite{tscherbul_partial_2015,tscherbul_long-lived_2014,dodin_quantum_2016} disappear if the incoherent radiation field is turned on adiabatically.
Similarly, these results clearly indicate that coherences observed in experiments utilizing fast laser pulses (e.g. \cite{collini_coherently_2010,engel_evidence_2007}) will not appear in nature where turn on times are essentially infinite on molecular time scales.}
Some explicit cases are discussed in Sec. VII below.


\section{Sample Light Harvesting Cases}
\label{sec:Examples}

The above results encompass a vast range of possible systems.
It is advantageous, therefore, to focus on some simple cases to emphasize the importance of these results to molecules of interest in, e.g., light harvesting scenarios. We address two sample questions below, being generous in our requirements for coherences. Note that we assume below that the system is isolated from an external (e.g. protein) environment, so as to focus solely on relaxation effects due to the incoherent light.  This artificial
arrangement is only designed to highlight some of the timescales
associated with the above analysis.

(i) Light-induced coherences have been observed experimentally in FMO, PC645 and other light harvesting complexes \cite{tiwari_electronic_2013,collini_coherently_2010,engel_evidence_2007}.
In these cases $\Delta \sim 100$ cm$^{-1}$ and $\gamma$, due to spontaneous emission, is on the order of $1$ ns$^{-1}$, which places the system in the overdamped region. For the sake of simplicity, this discussion neglects non-radiative relaxation and decay of the excited states due to the interaction with the phonons, which play an important role in realistic models of light-harvesting complexes \cite{tscherbul_donor-acceptor_????}.

Given our results above we can ask, for example: what turn-on time scales would be required to produce coherences that are even a modest $1\%$ of the population? Using \cref{eq:UDCRSl} shows that $|\rho_{e_1e_2}|/\rho_{e_ie_i}=\gamma/(\Delta_p^2\tau_r)$.
Hence, the turn-on time must be faster than $\sim 10-100$ ns, clearly far faster than natural turn-on times.
Hence, these coherences will not occur in natural light-harvesting systems.

(ii) Alternatively, we might ask what coherences (that are a modest $1\%$ of the population) can be generated by a turn-on time of $1$ ms, still a relatively fast turn-on time on natural time scales.

Here, using the same approach, we have $|\rho_{e_1e_2}|/\rho_{e_ie_i}=\gamma/(\Delta_p^2\tau_r)$.
Requiring this ratio to be a modest $1\%$ shows that states that will display coherences are separated by less than $0.9$ cm$^{-1}$.
Analogously, if we utilize a more realistic turn-on time of 1s, only levels separated by $9\times 10^{-4}$ cm$^{-1}$ will display coherences.
Once again, the results highlight the significance of the slow turn-on to assessing the (lack of) involvement of coherent phenomena in natural cases. 

\section{Conclusion}
\label{sec:Conc}
We have presented a generalization of the Bloch-Redfield master equations to the case of time-varying radiation fields.
They are shown to be of a similar structure to the previously studied master equations for stationary fields \cite{tscherbul_partial_2015,tscherbul_long-lived_2014,dodin_quantum_2016}, but with time-dependent incoherent pumping rates $r_i(t)$.
We explicitly determined the form of these master equations for the class of three-level V-systems and solved them analytically in the weak pumping limit relevant to the natural incident light (e.g. solar radiation).

Following the approach taken in the study of V-systems interacting with stationary fields \cite{dodin_quantum_2016,tscherbul_long-lived_2014}  two limiting cases  were considered in detail.
The underdamped regime ($\Delta_p \gg \gbar$) characterized by oscillatory coherences and the overdamped regime ($\gbar \gg \Delta_p$) characterized by quasistationary coherences.
In both regimes an inverse relationship between the maximal magnitude of the coherences and the turn on time $\max\{|\rho_{e_1e_2}|\} \propto 1/\tau_r$ in the adiabatic limit of very slow field turn-on was established.
This corresponds to a V-system in  equilibrium with the radiation field at all times.
In other words, the system is always approximately in the equilibrium mixture $\rho_{eq}=[\nbar(t),\nbar(t),0,0]^T$.

By contrast, for the very fast turn on of the radiation field, both regimes show dynamics that are identical to the sudden turn-on of the radiation field studied in the stationary field  case ($\tau_r \to 0$).
This limit occurs when the turn on time is much faster than any of the system timescales,
so that the system does not evolve under the transient field.
Instead it evolves under the steady state field that is reached very quickly.

For intermediate turn on times, the dynamics of the system can vary from those observed in the stationary field case but they, in general, reach the same maximal coherence as in the sudden turn on case.
One unexpected phenomenon observed was the synchronized decay of the coherences where all coherent superpositions decayed at the same time.
This differs from the naive expectation that coherent superpositions produced at later times would decay later than those produced at earlier times.
This synchronized decay of coherent superpositions occurs due to the suppression of excited state coherences by excited state populations.
When the coherences produced at early times decay, they lead to an increased population of the excited state manifold.
This subsequently leads to an increase in the decay rate of the coherent superpositions, which leads to a further increase in the excited state population.
Ultimately, this process leads to the run-away increase of the decay rate of the coherences at the decay time of the first superpositions prepared by the incident field and hence the synchronized decay of coherences.

These results reveal nontrivial effects of the turn on rate of the incoherent field on the dynamics of the system.
Most significantly they suggest that the significant coherences observed in the study of the V-system do not survive the slow turn-on of the radiation field.
Moreover, in the isolated molecule case, they will not survive for a field with a turn on time slower than the radiative lifetime of the excited states $\tau_{\gamma_i}=1/\gamma_i$ for a V-system in the underdamped limit or slower than the long time scale $\tau_\Delta=2\gbar/\Delta_p^2$ in the overdamped limit.
This greatly restricts the class of systems that would display significant coherences for radiation fields with physical turn-on times.

The implication of these results for pulsed laser experiments \cite{engel_evidence_2007,collini_coherently_2010,chenu_coherence_2015} that display coherences in biological molecules is profound.  Specifically, they imply that illumination by natural sunlight, where turn-on times are indeed enormously longer than all other relevant dynamical time scales, can not generate Fano coherences between other than essentially degenerate states.

{\bf Acknowledgements}

This work was supported by the US AFOSR through Contract No.
FA9550-13-1-0005, and by NSERC.

\appendix
\section{Generality of Results}
\label{sec:Generality}
To prove the generality of the results for the exponential turn on function \cref{eq:nt} presented in the main text, consider the set, $S$, of all continuous driving  functions, $\nbar(t)$, such that the function is initially zero and evolves to a steady state value, $\nbar$, in the long time limit.
That is
\beq
S=\left\{\nbar(t):[0,\infty)\to \mathbb{R}\; |\:\nbar(t)\in C_1; \; \nbar(0)=0;\; \lim_{t\to\infty}\nbar(t)=\nbar \right\}
\label{eq:nset}
\eeq
An element of $S$ in \cref{eq:nset} can, in general, be written as
\beq
\nbar(t)=\nbar-{g}(t)
\label{eq:nform}
\eeq
where $g(t) \in C_0([0,\infty), \mathbb{R})$ and $C_0([0,\infty), \mathbb{R})$ is the set of continuous functions from the interval $[0,\infty)$ on the real line to $\mathbb{R}$ which vanish at infinity.

We proceed now to prove that any function $g(t) \in C_0([0,\infty), \mathbb{R})$ can be written as a series of decaying exponentials on the positive real half-line.
This can be done using the Stone-Weierstrass theorem on locally compact spaces \cite{willard_general_2004}.
A set of functions, $A$, on $X$ is said to vanish nowhere if, for any $x \in X$, there exists a function, $f \in A$ such that $f(x) \neq 0$.
It is said to separate points if $\forall x\neq y \in X$ there exists a function $g \in A$ such that $g(x) \neq g(y)$.
Further, $C_0(X,\mathbb{R})$ defines an algebra over $\mathbb{R}$ under pointwise addition and multiplication of functions.

\begin{mythm} \textbf{(Stone-Weierstrass)}
Suppose $X$ is a locally compact Hausdorff space and $A$ is a subalgebra of $C_0(X,\mathbb{R})$. Then $A$ is dense in $C_0(X,\mathbb{R})$ if and only if it separates points and vanishes nowhere.
\label{thm:SW}
\end{mythm}

Begin by considering the interval $X=[0,\infty)$.
This is a closed subset of the locally compact Hausdorff space $\mathbb{R}$ and so is itself a locally compact Hausdorff space.
Define A as follows
\beq
A=span\{h_a(t)=e^{-a t}|a \in \mathbb{R}_+; \; t\in[0,\infty)\}
\label{eq:Aset}
\eeq
where $\mathbb{R}_+$ is the set of positive real numbers.
Clearly $A$ defines a vector space over the real numbers under pointwise addition and scalar multiplication of functions.
Furthermore, all decaying exponentials vanish at infinity so $A$ is contained in $C_0([0.\infty),\mathbb{R})$.
The product of linear combinations of decaying exponentials produces another such linear combination of exponentials, guaranteeing closure of $A$ under pointwise multiplication.
Therefore $A$ defines a subalgebra of $C_0([0,\infty),\mathbb{R})$.
It is trivial to show that $A$ vanishes nowhere and separates points on $[0,\infty)$.

Hence, according to this theorem, $A$ is dense in $C_0([0,\infty),\mathbb{R})$.
By the definition of a dense space, any function $g(t) \in C_0([0,\infty),\mathbb{R})$ is either in $A$ or is a limit point of $A$ \cite{steen_counterexamples_1995}.
In other words, any function, $g(t)$, vanishing at infinity on the positive real half-line can be expressed in the following form:

\beq
g(t)=\int_0^\infty da f(a) e^{-a t} =-\left(\int_0^\infty da f(a)(1-e^{-a t})\right)+\int^\infty_0 da f(a)
\label{eq:gform}
\eeq

Substituting \cref{eq:gform} into \cref{eq:nform}, and applying the initial condition $\nbar(0) =0$ yields the constraint $\nbar=\int^\infty_0daf(a) $.
This allows \cref{eq:nform} to be rewritten as
\beq
\nbar(t)=\int^\infty_0 da f(a)(1-e^{-a t})
\label{eq:nsum}
\eeq
\Cref{eq:nsum} expresses a general class of driving function as a series of terms each of the form considered in the main text \cref{eq:nt}.
The integral transform in \cref{eq:gform} is very similar to a Laplace Transform with the transformed coordinate, $a$, restricted to the real line rather than the complex plane \cite{boyce_elementary_2009}.
This yields an intuitive expansion of the time-dependent occupation number \cref{eq:nsum} in a basis where each basis function ($f_\alpha(t)=1-e^{-\alpha t}$) is associated with a characteristic turn on time $\tau_\alpha = 1/\alpha$.

Using \cref{eq:nsum}, the driving vector $\bm{d}(t)$ in \cref{eq:dvec} for an arbitrary driving function is given by
\beq
\bm{d}(t)= \left(\begin{array}{c}
\gamma_1 \\ \gamma_2 \\ p\sqrt{\gamma_1\gamma_2(t)}\\0
\end{array} \right)\int^\infty_0 daf(a)(1-e^{-a t})=\mathbf{d}\int^\infty_0 da f(a)(1-e^{-a t})
\label{eq:dt}
\eeq
Substituting \cref{eq:dt} into the general variation of parameters solution \cref{eq:GenSol} yields the solution for an arbitrary turn on function in terms of the solutions derived in the text.
\beq
\bm{x}(t)=\int^\infty_0 da f(a) \int^t_0 ds e^{A^{(0)}(t-s)}\mathbf{d}(1-e^{-a t})=\int^\infty_0 da f(a) \bm{x_a}(t)
\label{eq:AGenSol}
\eeq
where $\bm{x_a}(t)$ is the solution for a turn on function $\nbar_a(t)=(1-e^{-a t})$.

\Cref{eq:AGenSol} applied to the coherences indicates that any coherences observed are a result of the components with a fast turn on time.

%
%
\bibliography{vSysSlow}
\end{document}